\documentclass{article}
\usepackage{lajolla2006,graphicx}
\frompage{000} \topage{000}                                              
\def\rightmark{Viscosity at RHIC}
\def\leftmark{S. Pratt}
 
\title{Viscosity at RHIC} 

\authors{
{Scott Pratt and Kerstin Paech}\\[2.812mm]
{\normalsize
\hspace*{-8pt} Department of Physics and Astronomy, \\ 
\hspace*{-8pt} Michigan State University, East Lansing, MI~~48824-1321, USA 
}
}

\abstract{A variety of physical phenomena can lead to viscous effects. In this talk we review several sources of shear and bulk viscosity with an emphasis on the bulk viscosity associated with chiral restoration. We show that in the limit of a second order phase transition, the viscosity peaks in a singularity at the critical point.}

 
\begin{document}

\maketitle
\setcounter{page}{1}

\section{Introduction and Theory}
\label{sec:intro}
Viscosity has attracted remarkable attention at RHIC during the first years of running. In particular, experimental observations of large elliptic flow have pointed to a small shear viscosity and inspired the term ``perfect liquid'' \cite{RHIC_Whitepapers}. In this talk, we review the general theoretical definition of viscosity, then show how five different physical effects can lead to non-zero viscous coefficients (Sec. 2). We focus on bulk viscosity, and show that one can find large, even singular, effects in the neighborhood of $T_c$ (Sec. 3). Although the present study focuses on understanding the behavior and physical explanation of the coefficients, we speculate on the experimental manifestations large viscosities might bring about.

In non-viscous hydrodynamics the elements of the stress-energy tensor depend only on the energy density $\epsilon$ and particle-number densities $\vec{n}$ when viewed in the rest frame of the matter.
\begin{equation}
\tilde{T}_{ij}^{\rm (non.visc.)}({\bf r},t)
= \delta_{ij}P(\epsilon({\bf r},t),\vec{n}({\bf r},t)),
\end{equation}
where $\epsilon$ and $\vec{n}$ are implicitly functions of ${\bf r}$ and $t$. The tilde denotes that $T_{ij}$ is evaluated in a frame where the collective velocity ${\bf u}({\bf r})=0$. In Navier-Stokes hydrodynamics, viscosity is incorporated by altering $\tilde{T}$ so that it includes terms proportional to the velocity gradients $\partial u_i/\partial r_j$.
\begin{equation}
\label{eq:NS}
\tilde{T}_{ij}^{\rm (N.S.)} = \delta_{ij}P(\epsilon,\vec{n})
+\eta(\epsilon,\vec{n})\left(\frac{\partial u_i}{\partial r_j}+
\frac{\partial u_j}{\partial r_i}-\frac{2}{3}\nabla\cdot{\bf u}\delta_{ij}\right)
+B(\epsilon,\vec{n})(\nabla\cdot{\bf u})\delta_{ij},
\end{equation}
Here, $\eta$ and $B$ are the shear and bulk viscosities. Since $\nabla\cdot {\bf u}=(\partial\epsilon/\partial t)/(P+\epsilon)$, the bulk viscosity can be interpreted as a correction to the pressure due to the changing energy density, whereas the shear viscosity describes the asymmetry of $\tilde{T}_{ij}$ due to an anisotropic expansion. In non-viscous hydrodynamics accelerations are proportional to the gradient of the pressure, while in general, accelerations arise from derivatives of the stress energy tensor,
\begin{equation}
(\epsilon\delta_{ij}+\tilde{T}_{ij})\frac{\partial u_j}{\partial t}=
-\frac{\partial}{\partial x_j}\tilde{T}_{ij}.
\end{equation}
Thus, the components of the stress-energy tensor can be considered as representatives of the pressure in a given direction, and any reduction/rise of $\tilde{T}_{ij}$ from viscous effects will result in a slowing/acceleration of the expansion in that direction.

Viscous coefficients can be expressed in terms of correlations in the stress-energy tensor through Kubo relations. These are derived by considering alterations of $T_{ij}$ due to a perturbation $V$ in linear response theory,
\begin{equation}
\langle\delta T_{ij}(r=0)\rangle
=-(i/\hbar)\int_{r'_0<0} d^4r' 
\langle[T_{ij}(r=0),V(r')]\rangle,~~~
V(r')=r'_i(\partial_iu_j) T_{0j}(r'),
\end{equation}
where the perturbation represents the change to the Hamiltonian due to boosting according to a linear velocity gradient. After integrating by parts and applying the conservation of the stress energy tensor, $\partial_tT_{i,0}=-\partial_jT_{ij}$, one can derive the Kubo relation,
\begin{equation}
\label{eq:deltij}
\delta\langle T_{ij}(r=0)\rangle
=(i/\hbar)\int_{r'_0<0} d^4r' r'_0
\langle [\Delta T_{ij}(r=0),\Delta T_{kl}(r')]\rangle
\partial_k u_l.
\end{equation}
Here, $\Delta T_{kl}$ refers to the difference with respect to the average of $T_{kl}$ at $t=-\infty$. For $i\ne j$, symmetries constrain $k$ and $l$ to equal $i$ and $j$, which allows one to read off the shear viscosity from Eq. (\ref{eq:NS}),
\begin{eqnarray}
\label{eq:kuboshear}
\eta&=&(i/\hbar)\int_{r'_0<0} d^4r' r'_0
\langle[\Delta T_{ij}(0),\Delta T_{ij}(r')]\rangle,~~~i\ne j\\
\nonumber
&=&\lim_{\omega\rightarrow 0} \frac{-1}{2\omega\hbar}\int d^4r' e^{i\omega t'}
\langle[\Delta T_{ij}(0),\Delta T_{ij}(r')]\rangle.
\end{eqnarray}
By considering the case where $\partial_iu_j=(1/3)\delta_{ij} \nabla\cdot{\bf u}$, one can inspect $T_{ii}$ in Eq. (\ref{eq:deltij}) to find the bulk viscosity,
\begin{eqnarray}
\label{eq:kuboB}
B&=&(i/3\hbar)\sum_j \int_{r'_0<0} d^4r' r'_0
\langle [\Delta T_{ii}(0),\Delta T_{jj}(r')]\rangle\\
\nonumber
&=&\lim_{\omega\rightarrow 0} \frac{-1}{6\omega\hbar}\sum_j
\int d^4r' e^{i\omega t'}
\langle [\Delta T_{ii}(0),\Delta T_{jj}(r')]\rangle.
\end{eqnarray}
The Kubo relations, Eq.s (\ref{eq:kuboshear}) and (\ref{eq:kuboB}), are fully consistent with quantum mechanics. The classical limit can be obtained by applying the identity \cite{foerster},
\begin{equation}
e^{-\beta H}V(t)=e^{i\beta\hbar\partial_t}V(t)e^{-\beta H},
\end{equation}
to one of the terms in the commutator in Eq.s (\ref{eq:kuboshear}) or (\ref{eq:kuboB}), then keeping the lowest term in $\hbar$,
\begin{equation}
{\rm Tr}~ e^{-\beta H}[\Delta T_{ij}(0),\Delta T_{kl}(r)]
\approx -i\hbar\beta{\rm Tr}~
\partial_t \left(\Delta T_{ij}(0)Delta T_{kl}(r)\right),\\
\end{equation}
which after an integration by parts gives the classical limit of the Kubo relations,
\begin{eqnarray}
\eta&\approx&\beta\int_{r'_0<0} d^4r'
\langle\Delta T_{ij}(0)\Delta T_{ij}(r')\rangle,~~~i\ne j\\
B&\approx&(\eta/3)\sum_j \int_{r'_0<0} d^4r'
\langle \Delta T_{ii}(0)\Delta T_{jj}(r')\rangle
\end{eqnarray}
Although the Kubo relations are difficult to interpret physically, they do make it clear that viscosity is related to the size and to the damping of fluctuations of the elements $T_{ij}$. If fluctuations in $T_{ij}$ (at fixed energy) are large, or if they are slow to relax, a large viscosity will ensue.

\section{Five Sources of Viscosity}

Viscous effects arise whenever the elements of the stress-energy tensor, $T_{ij}$, have difficulty maintaining the equilibrium values due to a dynamically changing system, i.e., one with velocity gradients. In this section we briefly review five physical sources of viscosity, the first three of which have already been explained in the literature.

\begin{enumerate}
\item {\bf Viscosity from non-zero mean free paths}: This is the most commonly understood source of viscosity. It is straight-forward to see how a non-zero collision time leads to an anisotropy for $T_{ij}$ by considering a velocity gradient for a Bjorken expansion, $u_z=z/\tau$, i.e., $\partial_zu_z=1/\tau, \partial_xu_x=\partial_yu_y=0$. We consider a particle of momentum whose momentum is $p'_z(\tau)$ when measured in the frame moving with the collective velocity corresponding to its position. In the absence of collisions, $p'_z$ will fall with $\tau$ since the particle will asymptotically approach a region where its velocity equals the collective velocity, $p'_z(\tau+\delta\tau)
=p'_z(\tau)\tau/(\tau+\delta\tau)$. Meanwhile, $p'_x$ and $p'_y$ are frozen. The resulting anisotropy in the stress-energy tensor is easy to derive, which yields the following expression for the shear viscosity \cite{weinberg},
\begin{equation}
\eta=(4/3)P\tau_c,
\end{equation}
where $\tau_c$ is the collision time. It is also easy to see how such an expansion does not yield a bulk viscosity for either ultra-relativistic or non-relativistic gases. In those cases an isotropic expansion scales all three momenta proportional to $1/\tau$ which maintains thermal equilibrium, and collisions do not play a role. This is not the case when $m\sim T$, or especially if the gas has a mixture of relativistic and non-relativistic particles.

\item {\bf Viscosity from non-zero interaction range}: If the range of interaction between two-particles extends a distance $R$, interactions will share energy between particles from regions with different collective energies. A particle at $r=0$, where the collective energy is zero, will share energy with particles whose collective energy is $(1/2)m(R\partial_r u)^2$. For Boltzmann calculations, the viscosity will be proportional to $P R^2/\tau_c$ \cite{Cheng:2001dz}, with the constant of proportionality depending on the scattering kernel. Both bulk and shear terms result from non-zero interaction range. In Boltzmann calculations, the range of the interaction can approach zero for fixed scattering rates if the over-sampling ratio is allowed to approach infinity. Although this solves causality problems \cite{Kortemeyer:1995di}, it simultaneously eliminates viscous terms arising from finite-range scattering kernels, which might or might not be desirable. This has profound effects on calculations of elliptic flow, which can vary by a factor of 2 depending on the range of the scattering kernel \cite{Cheng:2001dz}.  

\item {\bf Classical Electric Fields}: Color flux tubes form after the exchange of soft gluons between nucleons passing at high energy, and might also be formed during rapid hadronization. Additionally, longitudinal color electric fields might be created during the pre-thermalized stage of the collision (color-glass condensate). Since these fields tend to align with the velocity gradient, they can be a natural source of shear viscosity. In fact, if the fields are purely longitudinal, the elements of the stress-energy tensor become $T_{zz}=-\epsilon, T_{xx}=T_{yy}=\epsilon$. Thus, the transverse pressure becomes three times as stiff as a massless gas, $P=\epsilon/3$, which is usually considered a stiff equation of state. The negative longitudinal pressure signifies that the energy within a given unit of rapidity is increasing during the expansion, similar to the negative work associated with stretching a rubber band. 

\item {\bf Non-equilibrium chemistry}: Chemical abundances can not keep up with the expansion if the rate at which equilibrium abundances change is not much smaller than the chemical equilibration rate, $1/\tau_{\rm chem}$,
\begin{equation}
\frac{dN}{dt}=-(1/\tau_{\rm chem})(N-N_{\rm eq}).
\end{equation}
If the equilbrium number is slowly changing, the abundances will vary from equilbrium by an amount,
\begin{equation}
\delta N=-\tau_{\rm chem}\frac{dN_{\rm eq}}{dt}.
\end{equation}
To associate this departure from equilibrium as a viscosity, one must consider the corresponding change in the pressure,
\begin{equation}
\delta P=\left.\frac{\partial P}{\partial n}\right|_{{\rm fixed~}\epsilon}\delta n,~~~
\frac{dN_{\rm eq}}{dt}=-\frac{\partial n}{\partial s}s\nabla\cdot{\bf u},
\end{equation}
where the second relation exploits the fact that entropy is conserved in a slow expansion. The bulk viscosity is then,
\begin{equation}
B=\left.\frac{\partial P}{\partial n}\right|_{{\rm fixed~}\epsilon}
\frac{\partial n}{\partial s}s.
\end{equation}
The bulk viscosity will be large whenever the equilibrium number is rapidly changing, e.g., the temperatures are falling below the masses, or masses are rising due to restoring chiral symmetry. If the hydrodynamic equations explicitly treat particle numbers as current obeying chemical evolution rates, chemical non-equilibration would not need to be accounted for through viscous terms.

\item {\bf Viscosity from dynamic mean fields}:
Bosonic mean fields, such as the $\sigma$ field, obey the Klein-Gordon equation. For fluctuations of wave number $k\rightarrow 0$,
\begin{eqnarray}
\label{eq:kg}
\frac{\partial^2}{\partial t^2}\Delta\sigma(t)
&=&-(m_\sigma(T)^2+k^2)\Delta\sigma(t)-\Gamma\frac{\partial}{\partial t}
\Delta\sigma(t),\\
\nonumber
\Delta\sigma(t)&\equiv&\sigma(t)-\sigma_{\rm eq}(\epsilon),
\end{eqnarray}
where $\sigma_{\rm eq}(\epsilon)$ is the equilibrium value of the condensate which is non-zero for $k=0$. The value of $\sigma$ is determined by minimizing the free energy, while the mass is related to the curvature of the free energy near the minimum,
\begin{equation}
\frac{\partial}{\partial \sigma}F(\sigma,T)=0,~~~
m_{\sigma,{\rm eq}}^2(T)
=\frac{\partial^2}{\partial \sigma^2}F(\sigma_{\rm eq},T).
\end{equation}
One can see the equivalence of Eq. (\ref{eq:kg}) with the differential equation for the harmonic oscillator after performing the following substitutions,
\begin{equation}
k_{\rm h.o.}/m_{\rm h.o.}\rightarrow m^2_{\sigma},~~~~
\gamma_{\rm h.o.}/m_{\rm h.o.} \rightarrow \Gamma,
\end{equation}
where $\gamma_{\rm h.o.}$ is the drag coefficient for the harmonic oscillator, $k_{\rm h.o.}$ is the  spring constant and $m_{\rm h.o.}$ is the particle mass. For the harmonic oscillator, the mean value of the position $x$ is altered if the equilibrium position is moving. The amount of the change was consistent with the drag force $\gamma dx_{\rm eq}/dt$ being equal and opposite to the restoring force $k\delta x$. The corresponding result can be derived for the damped Klein-Gordon equation,
\begin{equation}
\delta x=-\frac{\gamma_{\rm h.o.}}{k_{\rm h.o.}}\frac{dx_{\rm eq}}{dt},~~~~
\delta \sigma=-\frac{\Gamma}{m^2_{\sigma}(T)}\frac{d\sigma_{\rm eq}}{dt},
\end{equation}
where $\delta\sigma$ is the mean offset from the equilibrium value. Thus, $m^2_\sigma$ determines the restoring force, while $\Gamma$ describes the drag. Finite-size effects could be estimated by replacing $m^2$ with $m^2+k^2$, where $k^2$ would be given by the finite size, $k\sim 1/L$. The resulting bulk viscosity is,
\begin{equation}
B=\left.\frac{\partial P}{\partial \sigma}\right|_{{\rm fixed~}\epsilon}
\frac{\Gamma}{m_\sigma^2}\frac{\partial \sigma_{\rm eq}}{\partial s}s.
\end{equation}
The bulk viscosity is then large for energy densities where $\sigma$ is rapidly varying, or for when $m_\sigma$ is small, i.e., the critical region. 

\end{enumerate}

\section{Bulk Viscosity in the Linear Sigma Model}

Both of the last two sources of viscosity described in the previous section can be of special importance during the chiral transition. First, since masses are changing suddenly near $T_c$, chemical abundances should easily stray from equilibrium. Secondly, the mean field, which is zero above $T_c$ suddenly changes, and given the small masses in this region large bulk viscosities are expected. 

\begin{figure}[htb]
\centerline{\includegraphics[width=8cm]{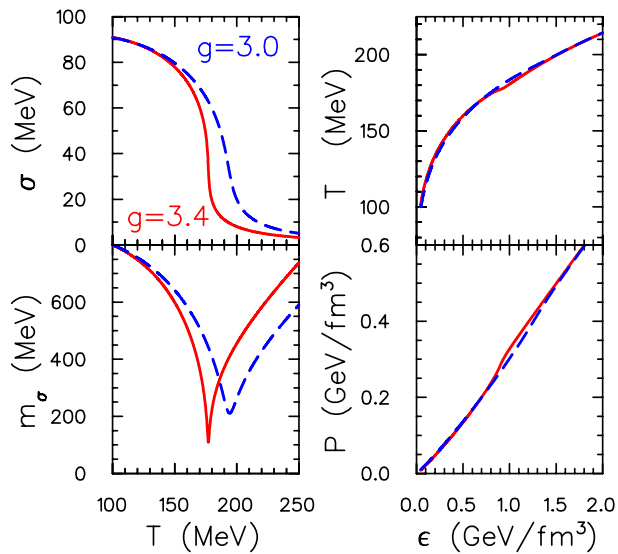}}
\caption{\label{fig:all4}For the linear sigma model, the sigma field and mass are shown as a function of the temperature in the left panels. Near $T_c$, the masses fall to zero and the mean value of the field changes rapidly, which gives rise to a sharp peak in the bulk viscosity. The pressure and temperature are displayed in the right-side panels as a function of energy density.}
\end{figure}
As an example, we consider a simple example of a linear sigma model, where the coupling of the sigma model to the quarks provides the quark mass \cite{Paech:2005cx,Paech:2003fe},
\begin{equation}
H=-\frac{1}{2}\sigma\nabla^2\sigma+\frac{\lambda^4}{4}\left(
\sigma^2-f_\pi^2+m_\pi^2/\lambda^2\right)^2-h_q\sigma
+H_{\rm quarks}(m=g\sigma),
\end{equation}
assuming only up and down flavored quarks. The resulting equation of state and values for $m_\sigma$ and $\sigma$ are displayed in Fig. \ref{fig:all4} for $\lambda^2=40$. For couplings $g<g_c=3.554$, the transition is a smooth cross-over, while for $g=g_c$ the transition is second order, and for $g>g_c$ a first-order phase transition ensues with $T_c=172$ MeV. From Fig. \ref{fig:all4}, one can see that $m_\sigma$ becomes small in the same region that the field rapidly changes, thus one expects a peak in the bulk viscosity. 

\begin{figure}
\centerline{\includegraphics[width=8cm]{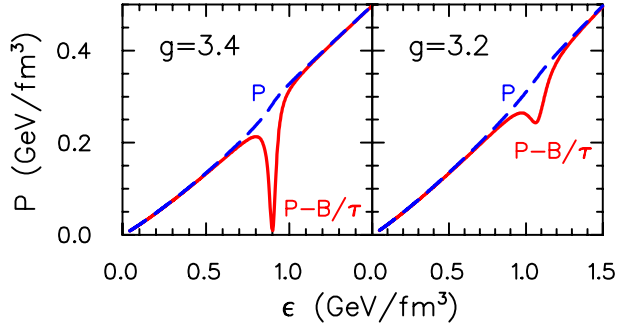}}
\caption{\label{fig:visctau}For a Bjorken expansion ($\nabla\cdot{\bf u}=1/\tau$), the pressure is plotted alongside $T_{ii}=P-B\nabla\cdot{\bf u}$, to demonstrate the significance of viscous terms near $T_c$. Viscous terms are larger and sharper for couplings close to the critical coupling.}
\end{figure}
The bulk viscosity was calculated according to the methods described in (4) and (5) of the previous section assuming that the width $\Gamma=400$ MeV, and that the chemical equilibration time $\tau_{\rm chem}$=1 fm/$c$ for a density of one quark per fm$^3$ and scaling inversely with the density. For a Bjorken expansion $\nabla\cdot{\bf u}=1/\tau$, and assuming an isentropic expansion starting with $\epsilon=8$ GeV/fm$^3$ at $\tau=1$ fm/$c$, we calculated both $P$ and $B$ as a function of $\tau$. To illustrate the size of the effect, we display both $P$ and the Navier-Stokes expression $T_{ii}=P-B\nabla\cdot{\bf u}$ as a function of the energy density for a Bjorken expansion in Fig. \ref{fig:visctau}. The effect is certainly dramatic, especially for $g\approx g_c$. However, the Navier-Stokes expression is only applicable for small expansion rates. We expect Israel-Stewart \cite{Israel:1979wp,Heinz:2005bw} equations for hydrodynamics  to result in moderated effects compared to Navier-Stokes, though they should give identical results for small velocity gradients. 

\section{Summary}

The simplicity of the Kubo relations, Eq.s (\ref{eq:kuboshear}) and Eq. (\ref{eq:kuboB}), masks the wide variety of physical sources of viscosity. The one common aspect of the various sources is that non-zero equilibration times, non-zero interaction ranges can always be identified. In this talk, we focused on bulk viscosities associated with the chiral transition. In general, one would expect such effects whenever a system needs to rapidly rearrange its basic structure. In this sense these effects have much in common with super-cooling or hysteresis.

The implications for dynamics should be that the matter accelerates more quickly due to the higher gradients in $T_{x,x}$ when the interior energy density is above the critical region. However, once the matter flows into the viscous region  of energy densities, there should be a slowing down and a reduction of surface emission. This trend would be in the right direction to explain HBT measurements which show a rapid expansion with a sudden disintegration \cite{Retiere:2003kf}, but the potential magnitude of the effects are not yet known.

Finally, we re-emphasize that if one were to solve for the evolution of the mean fields or chemistry alongside solving the hydrodynamic evolution equations, one might be able to neglect some of these effects. If these effects are large, the proper conclusion may be that rather than absorbing these effects into viscous hydrodynamics, one should treat non-equilibrated degrees of freedom more explicitly.

\section*{Acknowledgments}
Support was provided by the U.S. Department of Energy, Grant No. DE-FG02-03ER41259.

\end{document}